\documentclass[a4paper,11pt]{article}
\usepackage{pos}
\usepackage{subfigure}

\newcommand{\Nff}{N_f^{(F)}}
\newcommand{\Nfa}{N_f^{(A)}}

\title{One flavour adjoint QCD with overlap fermions}

\author*[a]{G.~Bergner}
\author[a]{I.~Soler Calero}
\author[a]{J.~C.~Lopez}
\author[b]{S.~Piemonte}

\affiliation[a]{University of Jena, Institute for Theoretical Physics, Max-Wien-Platz 1, 07743 Jena, Germany}
\affiliation[b]{Institute for Theoretical Physics, University of Regensburg, 93040 Regensburg, Germany}

\emailAdd{georg.bergner@uni-jena.de}

\abstract{The infrared effective theory of adjoint QCD with one Dirac flavour is still under debate. The theory could be confining, conformal, or with a massless fermion in the infrared. The study of chiral symmetry seems to be important to answer this question. While previous investigations have considered Wilson fermions, we present here the first results for this theory based on overlap fermions to avoid explicit chiral symmetry breaking. These indicate spontaneous chiral symmetry breaking by the formation of a fermion condensate. We have also investigated the running coupling of the theory, which indicates no infrared conformality in the energy region we have explored.}

\FullConference{%
The 39th International Symposium on Lattice Field Theory,\\
8th-13th August, 2022,\\
Rheinische Friedrich-Wilhelms-Universität Bonn, Bonn, Germany
}

\begin{document}
\maketitle

\section{Yang-Mills theory with one  adjoint flavor}
Yang-Mills theories coupled to fermions in the adjoint representation have received much interest as a realization of strong interactions that is quite different from standard QCD like theories with fermions in the fundamental representation \cite{universe}. With adjoint fields, an explicit breaking of center symmetry is avoided and the deconfinement transition can be identified in an unambiguous way. These theories are closely related to $\mathcal{N}=1$, $\mathcal{N}=2$, and $\mathcal{N}=4$ supersymmetric Yang-Mills theory (SYM) if scalar fields are neglected. 

We are denoting the number of adjoint Dirac fermion fields as $\Nfa$, which means that Majorana flavors are effectively equal to half a Dirac flavor and $\Nfa=\frac12$ adjoint QCD corresponds hence to $\mathcal{N}=1$ SYM. We have investigated the non-perturbative properties of $\mathcal{N}=1$ SYM in several studies, considering the gauge group SU(2) and SU(3) \cite{Bergner:2015adz,Ali:2019agk}. Another theory considered in several numerical investigations is $\Nfa=2$ SU(2) adjoint QCD since it has been considered as a possible composite Higgs extension of the Standard Model, see e.~g.\ \cite{Bergner:2016hip,DelDebbio:2009fd,Hietanen:2009az} and references therein. Theories with $\Nfa=\frac32$ \cite{Bergner:2017gzw} and $\Nfa=1$ have also been investigated. 

In particular, $\Nfa=1$ adjoint QCD has recently gained much interest. Different infrared scenarios have been suggested for this theory and numerical simulations might help to determine which of them is realized. A possible scenario is that the theory is confining similar to $\mathcal{N}=1$ SYM. In this case, chiral symmetry is spontaneously broken from $\text{SU}(2\Nfa)$ to $\text{SO}(2\Nfa)$ and massless Goldstone bosons appear in the infrared. The theory could also be inside the conformal window, which means that the infrared limit is determined by a conformal fixed point of the running gauge coupling. In this case masses of all particles scale to zero in the chiral limit. A third scenario has been conjectured based on 't Hooft anomaly matching \cite{Anber:2018iof}. The infrared effective theory is in this case determined by light fermion bound states. The investigation of these infrared properties of the theory has been the main motivation of several investigations with numerical methods of lattice field theory \cite{Athenodorou:2014eua,Bi:2019gle,Athenodorou:2021wom}. 

It is, however, hard to achieve a final conclusion for this theory since it is at least quite close to a conformal behavior. Large lattice volumes are required, but also possible bulk phases induced by a coarse lattice spacing have to be excluded. In order to identify spontaneous chiral symmetry breaking, an unambiguous order parameter would be desirable, which is absent in the standard Wilson fermion discretizations used in investigations so far.

Our study is a first consideration of an improved lattice discretization for the theory \cite{Bergner:2022hoo}. We have performed simulations with overlap fermions, which provide a well defined representation of chiral symmetry on the lattice and reduced lattice artefacts. Our investigations are part of a larger exploration of the theory space with $\Nff$ fermions in the fundamental and $\Nfa$ fermions in the adjoint representation. We have already explored theories including both, fundamental and adjoint fields, in our numerical simulations \cite{Bergner:2020mwl}. 

One main long term goal is to approach the supersymmetric limit in this theory space once also scalar fields are added. Since supersymmetry is broken on the lattice, a fine tuning of a larger number of counterterms is required. This tuning is significantly simplified with the overlap fermion formulation since chiral symmetry reduces the number of possible counterterms. The exploratory study with overlap fermions is hence also a first simplified step towards a simulation of supersymmetric gauge theories.

In this short presentation we first introduce the approximate overlap formulation that we have applied in our simulations. We have used the same setup in $\mathcal{N}=1$ SYM before and achieved results consistent with Domain-Wall fermion realizations and Gradient flow \cite{Piemonte:2020wkm}. One main result is the chiral condensate, which provides a direct evidence for chiral symmetry breaking. In order to provide more evidence concerning the existence of an IR fixed point of the theory, we also estimate the running coupling of one flavor adjoint QCD. 

\section{Overlap fermion approximation and simulation parameters}
The Ginsparg-Wilson relation provides a well defined chiral symmetry and an unambiguous chiral condensate on the lattice. This implies that the magnitude of the condensate can be in principle directly determined in the simulations, providing an answer to the question about the infrared effective theory of $\Nfa=1$ adjoint QCD. However, in practice, approximations and extrapolations are still required.

The implementation of chiral symmetry on the lattice in terms of a Ginsparg-Wilson relation always comes at an additional cost. The operator requires an approximation of the matrix valued sign function and a basic operator kernel, which is in our case the hermitian Dirac-Wilson operator, $D_{\text{H}}(\kappa)=\gamma_5 D_w(\kappa)$, depending on the Hopping parameter $\kappa$. The complete overlap operator is
\begin{align*}
D_{ov}=\frac{1}{2}+\frac{1}{2}~\gamma_5~\text{sign}(D_{\text{H}}(\kappa))\, ,\quad 
\text{sign}(D_{\text{H}})=\frac{D_{\text{H}}(\kappa)}{\sqrt{D_{\text{H}}(\kappa)~D_{\text{H}}(\kappa)}}\; .
\end{align*}
One can optimize the approximation of the sign function for a given range of the eigenvalue spectrum of $D_{\text{H}}$.
This can be done using, for example, rational, polynomial, or other approximations schemes. These approximations can be expressed also in a higher dimensional space like done for variants of the domain wall operator. 

However, the main challenge is not only to find an optimized approximation. The derivative of the sign function is required in the rational hybrid Monte Carlo algorithm (RHMC), and it diverges towards the origin. The approximation corresponds also to a regularization of the divergence in the force calculation. In our exploratory study, we have used a polynomial approximation (order $N$) of the inverse square root. This has the additional advantage that it can be easily integrated into the rational approximation of the RHMC. In practice, this means that the final results in the chiral limit are obtained from an extrapolation of $N\rightarrow \infty$. As explained below, we have observed that finite $N$ effectively implies a finite residual mass and a gap in the eigenvalue spectrum, i.~e.\ a non-zero effective lower bound for the real part of the eigenvalues of $D_{ov}$ considering the relevant configurations.

The regularization of the overlap operator leads also to meaningful values of the chiral condensate. Towards the chiral limit, this observable gets divergent contributions from a decreasing number of configurations. This implies a divergent error for the numerical value of the condensate. The extrapolation of $N\rightarrow \infty$ can, however, be done in a parameter space where the numerical uncertainty is still acceptable. 

\begin{figure}
	\subfigure[violation of Ginsparg-Wilson relation \label{fig:GWviolation}]
	{\includegraphics[width=0.51\textwidth]{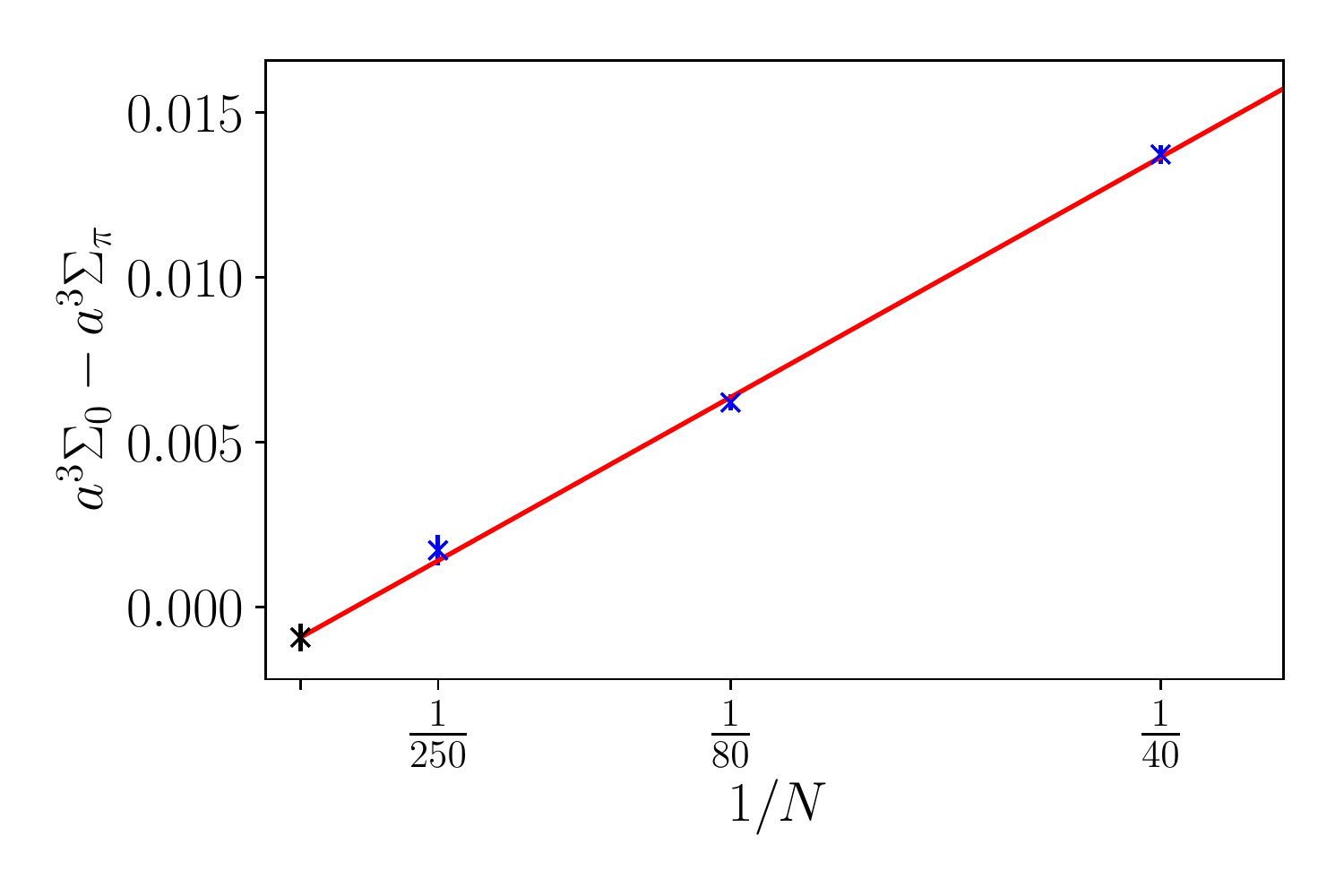}}
	\subfigure[spectrum of approximate overlap operator \label{fig:oveigenvalues}]
	{\includegraphics[width=0.43\textwidth]{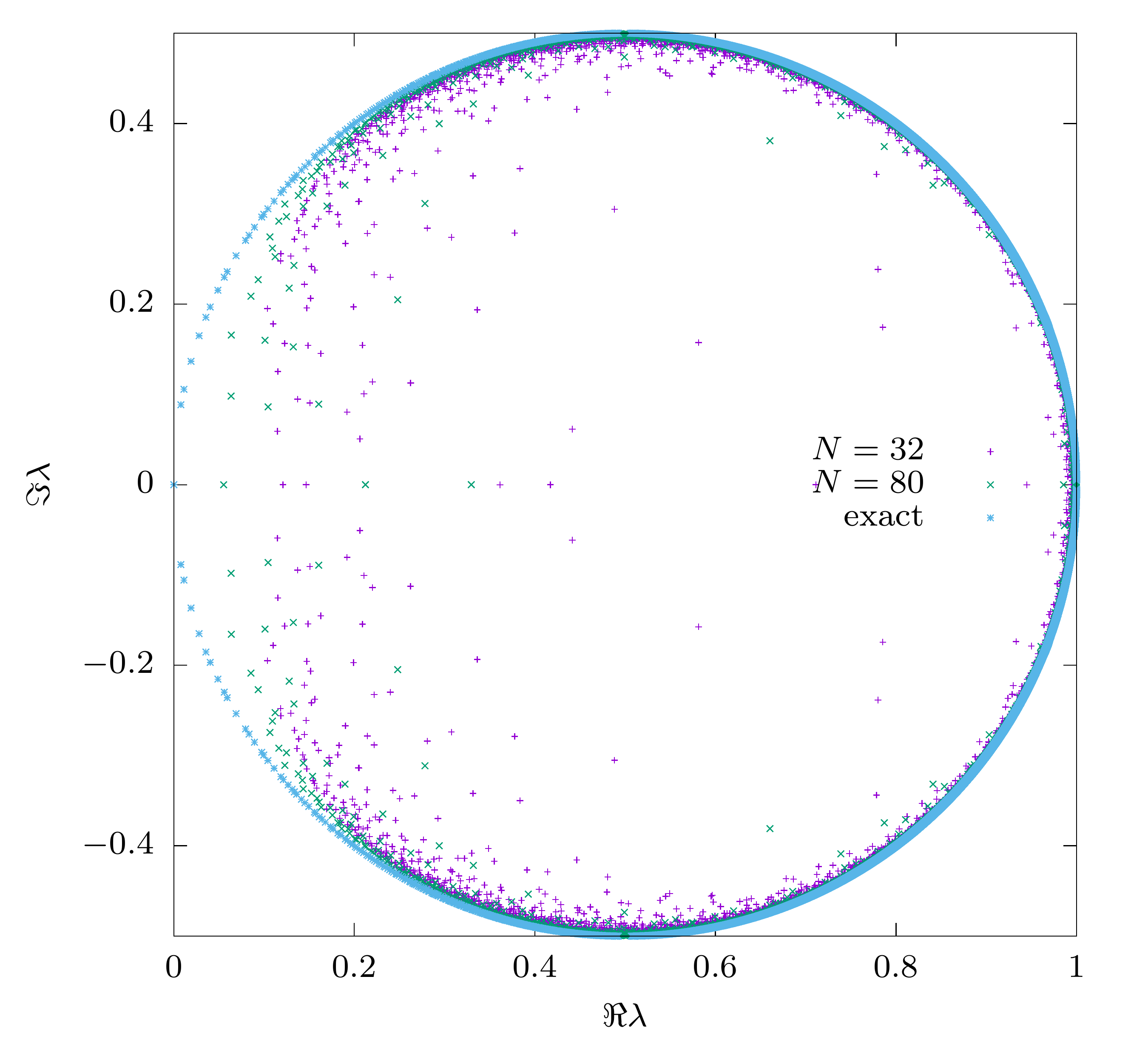}}
	\caption{Violation of the Ginsparg-Wilson relation at finite $N$ for our polynomial approximation. Left hand side: direct measurement of the chiral symmetry violation using the difference between the chiral condensate before and after rotation ($\beta=1.6$). Right hand side: complete eigenvalue spectrum of the overlap operator on an $8^4$ lattice. A comparison of the polynomial approximation at order $N$ and the exact overlap operator is shown. \label{fig:approx}}
\end{figure}
The effect of different $N$ in the polynomial approximation is shown in Fig.~\ref{fig:approx} using the difference between the expectation value of the condensate before and after a certain rotation based on the chiral symmetry encoded in the Ginsparg-Wilson relation. It basically measures the violation of the Ginsparg-Wilson relation, which can be interpreted as a residual mass term. We observe good agreement with a linear dependence on $1/N$, see Fig.~\ref{fig:GWviolation}. Hence we assume that the residual mass scales like $m_{res}\propto 1/N+O((1/N)^2)$ towards the chiral limit. The quality of the approximation can also be seen in Fig.~\ref{fig:oveigenvalues} showing the eigenvalue of the approximate overlap operators, which should approximate a circle in the large $N$ limit.

We applied a tree level Symanzik improved gauge action and additional stout smearing of the links inside $D_{ov}$. The inverse gauge coupling has been in the range $\beta=1.6-1.8$. The hopping parameter inside the kernel $D_\text{H}$ was chosen to be $\kappa=0.2$ according to the observed $D_w$ eigenvalue spectrum.

\section{Results from numerical simulations}
Our main result obtained with the overlap fermions is the chiral condensate providing evidence of a spontaneous chiral symmetry breaking, disfavoring a conformal scenario. In addition, we provided data for the running of the gauge coupling, which also shows no indications of a conformal fixed point. 
\subsection{Chiral condensate}
\begin{figure}
	\subfigure[$\beta=1.6$ \label{fig:chiralcond16}]
	{\includegraphics[width=0.49\textwidth]{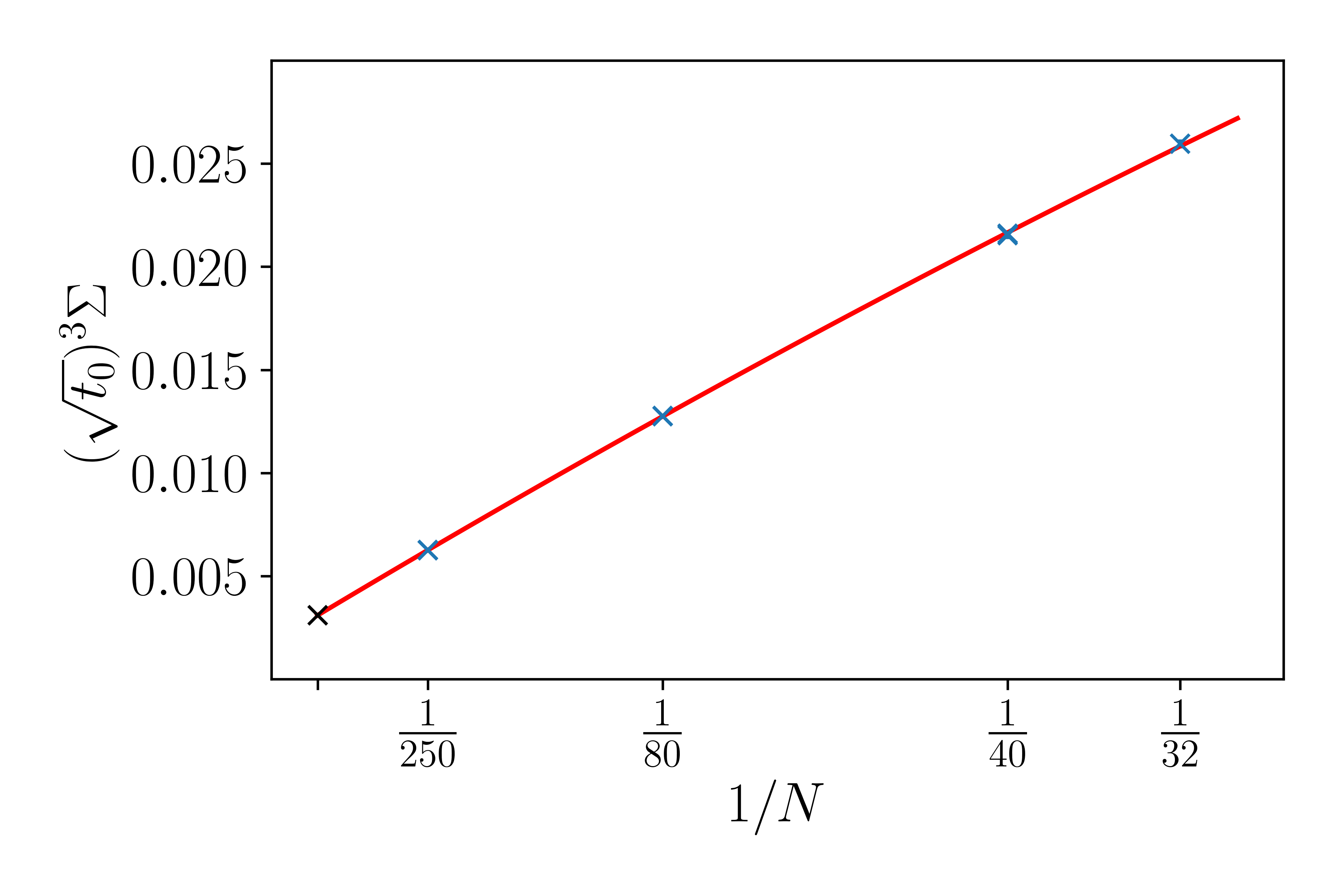}}
	\subfigure[$\beta=1.7$ \label{fig:chiralcond17}]
	{\includegraphics[width=0.49\textwidth]{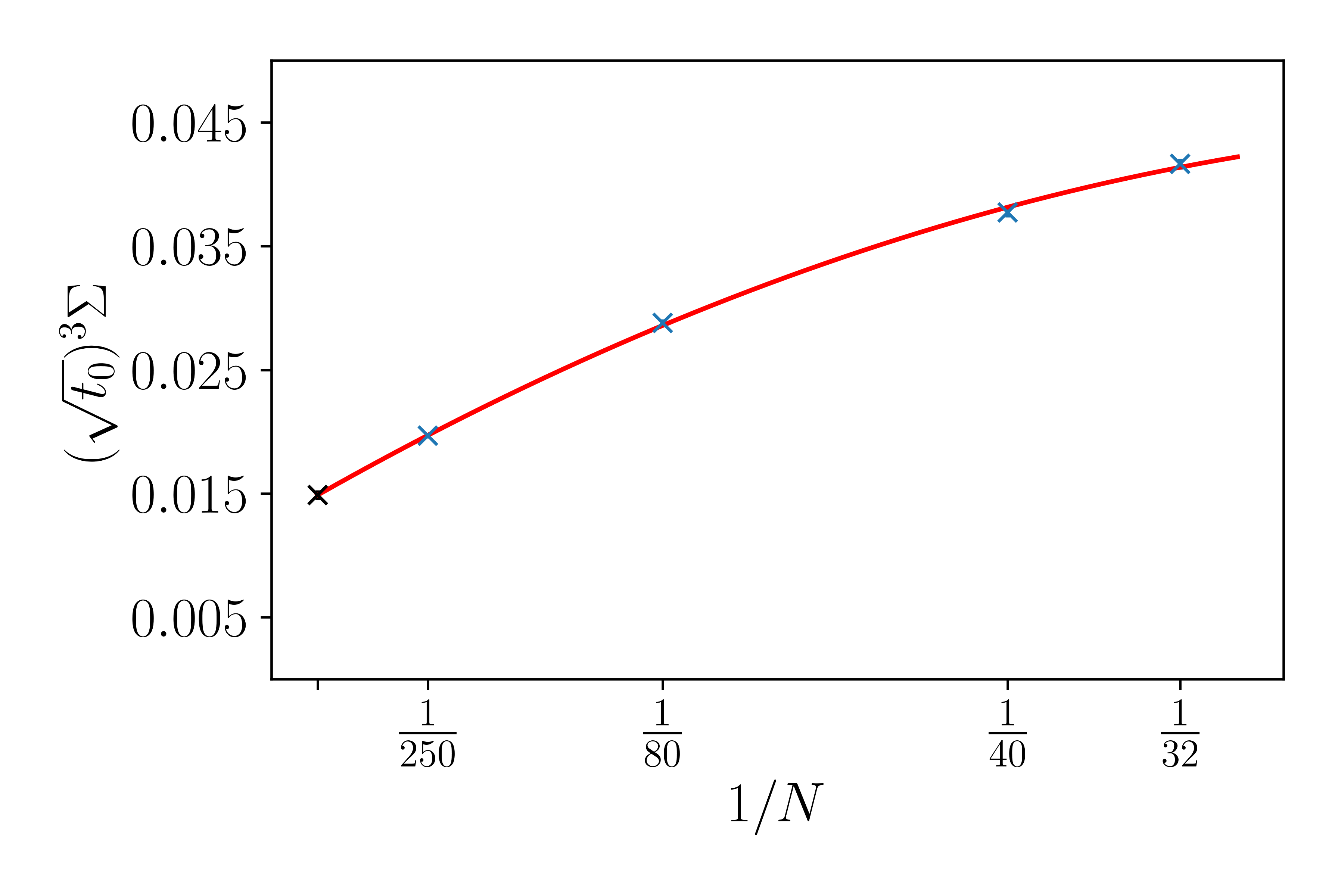}}
	\caption{Condensate as a function of the inverse order of the polynomial approximations for two different values of the inverse gauge coupling $\beta$. The condensate is given in units of the gradient flow scale $t_0$. An extrapolation to the infinite $N$ (chiral) limit is done using a linear fit with quadratic correction.  \label{fig:chiralcondensate}}
\end{figure}
The chiral condensate can be extrapolated to the infinite $N$ limit as shown in Fig.~\ref{fig:chiralcondensate}. As explained above, the dependence on $1/N$ can be translated into the one on the residual mass or the violation of the Ginsparg-Wilson relation. We observe good agreement using a linear dependence with quadratic corrections for the fit. The important result is that all fits indicate a finite value of the condensate in the chiral limit. This is a first hint for a chiral symmetry breaking disfavoring a conformal scenario for this theory.

\subsection{Running coupling from scale setting}
\begin{figure}
	\centerline{\includegraphics[width=0.6\textwidth]{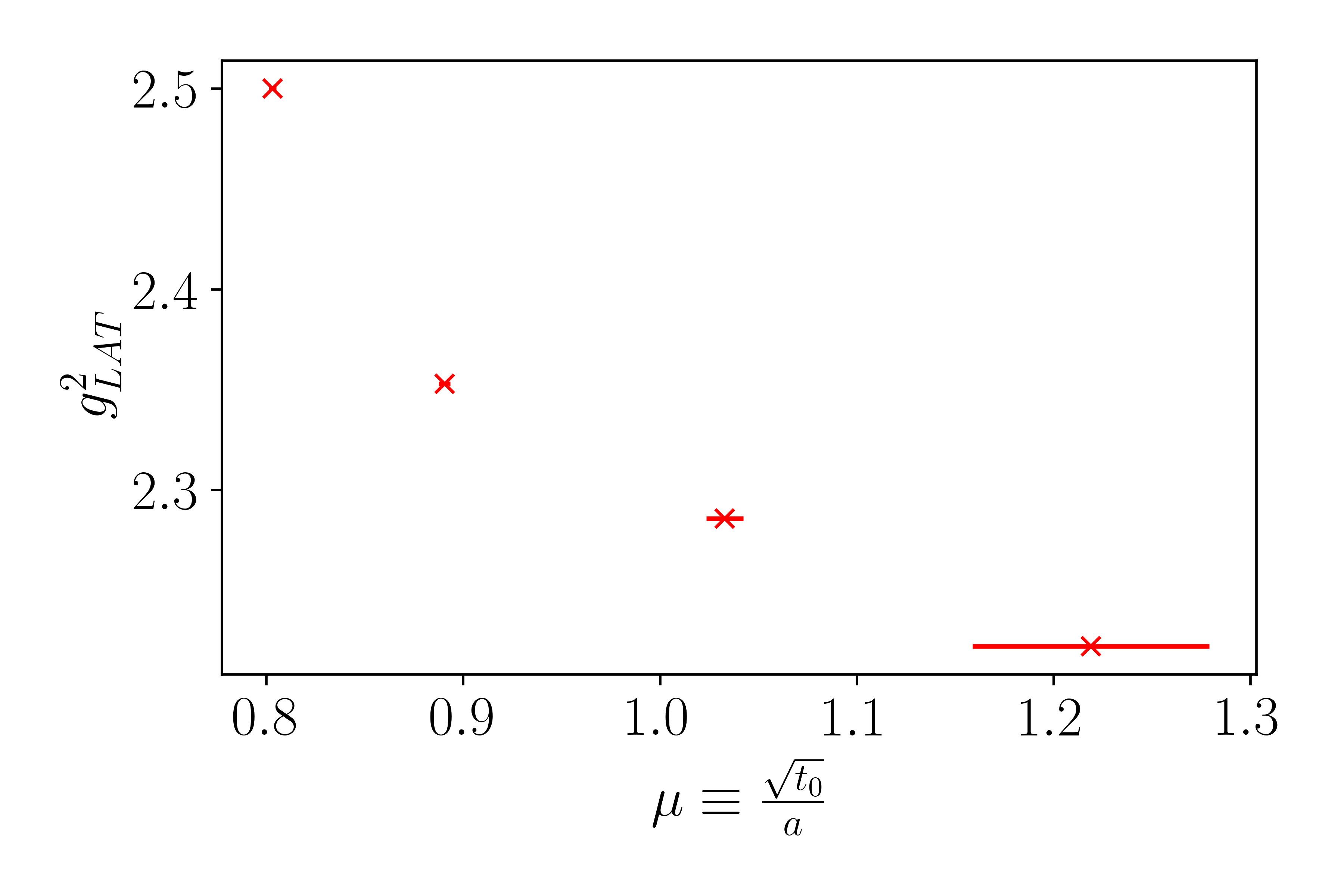}}
	\caption{Scale dependence of the strong coupling obtained from the lattice scale setting based on the gradient flow $t_0$. The four different points correspond to the four different bare gauge couplings used in the simulations.  \label{fig:barerunning}}
\end{figure}
The chiral condensate has provided a first evidence for a chiral symmetry breaking behavior. However, it does not show whether there is any indication for a conformal fixed point in the running of the strong coupling. In this first exploratory study, we are not able to perform a complete analysis of the beta function, but we can still obtain first estimates based on the scale setting at the different values of the inverse gauge coupling. We perform the scale setting in a mass independent way by extrapolating $t_0$ to the infinite $N$ (chiral) limit.

In the simplest approximation, the strong coupling is just given by the bare gauge coupling and the scale is determined by the inverse lattice spacing in units of $t_0$. The running strong coupling obtained in this simplistic approach is depicted in Fig.~\ref{fig:barerunning}. The general scale dependence is very similar to QCD and shows no indication of a fixed point.

\subsection{Running coupling from gradient flow}
\begin{figure}
	\centerline{\includegraphics[width=0.6\textwidth]{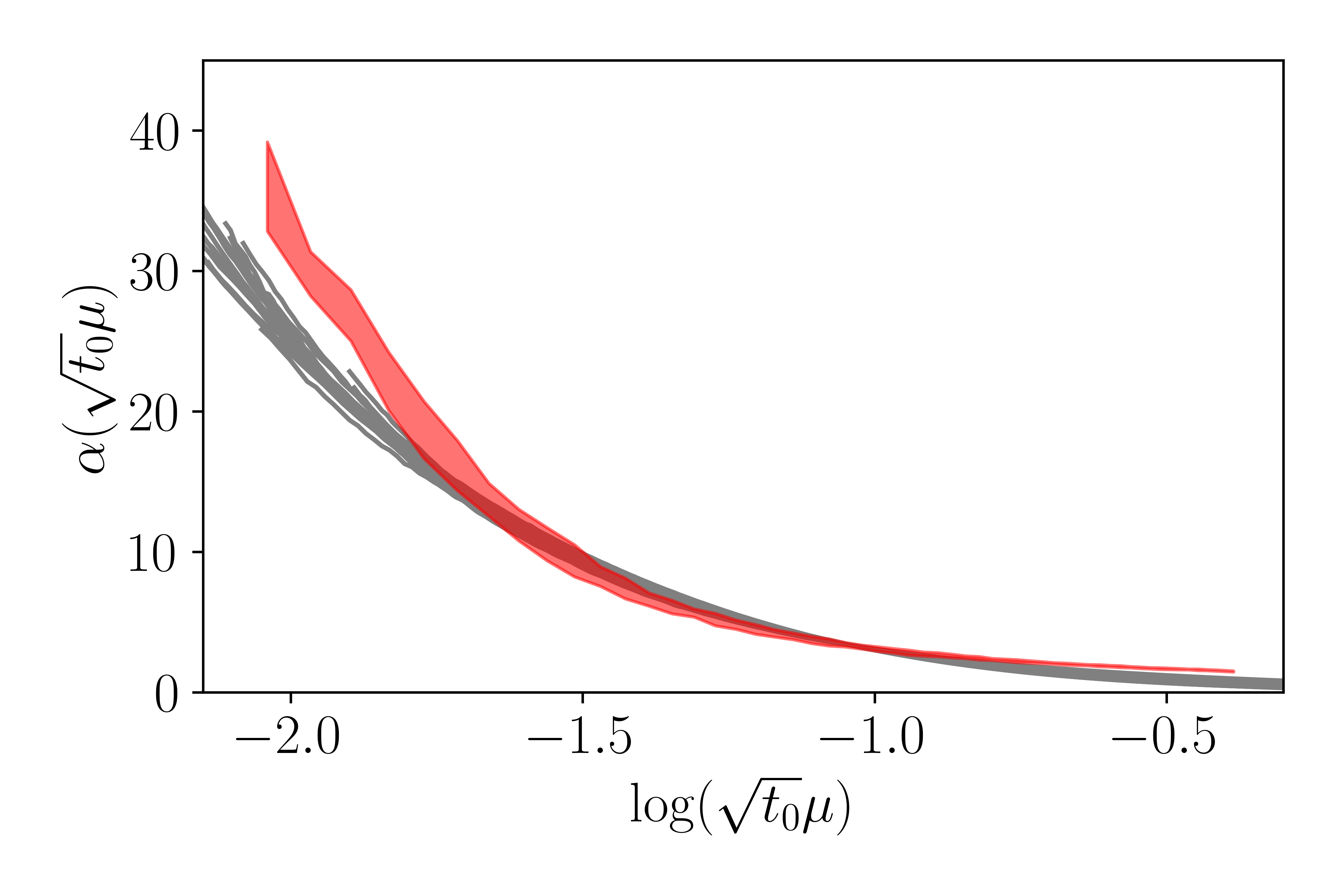}}
	\caption{Scale dependence of the gradient flow strong coupling defined in \eqref{eq:gfstrongcoupling}. The gray lines indicate the raw data and the red band corresponds to the extrapolation \eqref{eq:gfstrongcoupling_extr}.  \label{fig:contrunning}}
\end{figure}
A more precise and measurement of the scale dependent strong coupling can be obtained from the gradient flow. 
One can define a scale dependent gradient flow coupling from the dependence of the energy density $E$ on the flow time $\tau$
\begin{align}\label{eq:gfstrongcoupling}
g^2_{\textrm{GF}}(\mu)=\frac{16\pi^2}{3(N^2-1)}\tau^2\left.\langle E(\tau)\rangle \right|_{\tau^2=1/8\mu}\,.
\end{align}
The obtained running coupling needs to be extrapolated to the chiral and continuum limit. With the coarse lattice that we are considering here, finite volume corrections are not relevant. Hence we perform the following extrapolations of the coupling at a given scale $\mu$ (with undetermined coefficients $c_0$, $c_1$, $d_1$, $d_2$, and $d_3$):
\begin{align}\label{eq:gfstrongcoupling_extr}
g(\mu)^2 = g_0 + c_0 t_0^{-1} + c_1 t_0^{-\frac{3}{2}} + d_1 \frac{\sqrt{t_0}}{N} + d_2 \frac{1}{N} + d_3 \left(\frac{\sqrt{t_0}}{N}\right)^2\,.
\end{align}

The final results of this analysis shown in Fig.~\ref{fig:contrunning} confirm the rough estimates obtained from simple scale setting: the overall scale dependence seems to be QCD-like. There is no indication of a particularly small running or scale independent behavior at the parameters of our simulations.

\section{Summary and conclusions}
Adjoint QCD with one Dirac flavor is an interesting theory since it is connected to proposed extensions of the standard model, supersymmetric gauge theories, and conjectures about alternative infrared scenarios for strongly interacting gauge theories. It is probably quite close to the lower end of the conformal window, which makes it a more challenging target for numerical lattice simulations. For this reason, it remains still controversial what kind of infrared scenario is realized. To avoid possible ambiguities and deformations of the infrared effective theory, it is beneficial to preserve as much symmetries as possible on the lattice. Ginsparg-Wilson fermions are hence an ideal tool for such an investigation. They also provide a direct way to study chiral symmetry breaking.

Ginsparg-Wilson require, however, some regularized version of the sign function. In our first studies, we have used a rather simple approach with a polynomial approximation of order $N$. We have observed that reasonable extrapolations to the chiral (infinite $N$) limit are possible.

The results that we obtained in this approach all hint towards a chiral symmetry breaking QCD-like scenario. We obtain a finite value for the chiral condensate and an exponentially increasing coupling towards lower energies. There are no indications for a conformal scenario. 

It is quite clear that our current investigations have quite some limitations, especially considering the rather coarse lattice spacings. It is therefore important to extend the considered parameter range. This can also be done using complementary implementations of Ginsparg-Wilson fermions, like domain wall fermions. 

\noindent{\bf Acknowledgements:} G.~B.\ and I.~S.\ are funded by the Deutsche Forschungsgemeinschaft (DFG) under Grant No.~432299911 and 431842497.
The authors gratefully acknowledge the Gauss Centre for Supercomputing e.~V.\ (www.gauss-centre.eu) for funding this project by providing computing time on the GCS Supercomputer SuperMUC-NG at Leibniz Supercomputing Centre (www.lrz.de). Further computing time has been provided on the compute cluster PALMA
of the University of M\"unster and resources of Friedrich Schiller University Jena supported in part by DFG grants INST 275/334-1 FUGG and INST 275/363-1 FUGG. 


\end{document}